%% file: main.tex
\documentclass[10pt,journal]{IEEEtran}

\usepackage{siunitx} 
\usepackage[final]{graphicx} 
\usepackage{amsmath} 
\usepackage{amsthm} 
\usepackage{amssymb,times}
\usepackage{subfigure}
\usepackage{cite}
\usepackage{algorithm}
\usepackage{algorithmic}
\usepackage[english]{babel}
\usepackage{color}
\usepackage{url}
\usepackage{cases}
\usepackage{hyperref}
\input{common}

\begin{document}
\title{Applications of Absorptive Reconfigurable Intelligent Surfaces in Interference Mitigation and Physical Layer Security
}

\author{Fangzhou Wang,~\IEEEmembership{Member,~IEEE,} and
 A. Lee Swindlehurst,~\IEEEmembership{Fellow,~IEEE,}
\thanks{F. Wang and A. Lee Swindlehurst are with the Center for Pervasive Communications and Computing, Henry Samueli School of Engineering, University of California, Irvine, CA 92697 USA (e-mail: Fangzhou.Wang@aptiv.com; swindle@uci.edu)}
}

\maketitle

\begin{abstract}
This paper explores the use of reconfigurable intelligent surfaces (RIS) in mitigating cross-system interference in spectrum sharing and secure wireless applications. Unlike conventional RIS that can only adjust the phase of the incoming signal and essentially reflect all impinging energy, or active RIS, which also amplify the reflected signal at the cost of significantly higher complexity, noise, and power consumption, an absorptive RIS (ARIS) is considered. An ARIS can in principle modify both the phase and modulus of the impinging signal by absorbing a portion of the signal energy, providing a compromise between its conventional and active counterparts in terms of complexity, power consumption, and degrees of freedom (DoFs). We first use a toy example to illustrate the benefit of ARIS, and then we consider three applications: (1) Spectral coexistence of radar and communication systems, where a convex optimization problem is formulated to minimize the Frobenius norm of the channel matrix from the communication base station to the radar receiver; (2) Spectrum sharing in device-to-device (D2D) communications, where a max-min scheme that maximizes the worst-case signal-to-interference-plus-noise ratio (SINR) among the D2D links is developed and then solved via fractional programming; (3) The physical layer security of a downlink communication system, where the secrecy rate is maximized and the resulting nonconvex problem is solved by a fractional programming algorithm together with a sequential convex relaxation procedure. Numerical results are then presented to show the significant benefit of ARIS in these applications.
\end{abstract}

\begin{IEEEkeywords}

Reconfigurable intelligent surfaces, interference mitigation, spectral coexistence, device-to-device communication, physical layer security.
\end{IEEEkeywords}

\IEEEpeerreviewmaketitle

\section{Introduction}
Reconfigurable Intelligent Surfaces (RIS) can configure the wireless propagation environment by tuning the properties of individual low-cost (essentially) passive elements, and have attracted significant attention within both the radar and wireless communication communities \cite{HuangAlexandropoulos2019TWC,PanRenSwindlehurst2021MCOM,ZhangDaiPoor2021arXiv,JianAlexandropoulos2022arXiv,ZhangZhangSong2022TWC,BuzziGrossiLops2022TSP}. Some recent works on RIS include joint active and passive beamforming design \cite{WangFangLi2021TWC}, channel estimation \cite{SwindlehurstZhouLiu2022JPROC}, reflection modulation analysis \cite{ShaoeZhengAlexandropoulos2021TWC}, compressive sensing and deep-learning-based solutions \cite{TahaAlrabeiah2021ACCESS}, holographic designs \cite{LiAlexandropolos2022JSTSP,LiHuangAlexandropoulos2023TWC,gong2023holographic}, along with many other interesting application areas. In the radar community, RIS are employed to achieve non-line of sight surveillance \cite{AubryMaio2021TVT}, enhance the system reliability for target detection \cite{BuzziGrossi2021SPL}, improve parameter estimation performance \cite{EsmaeilbeigMishra2021arXiv}, and suppress clutter interferences \cite{WangLiFang2022SPL}. In addition, RIS have recently been shown to play an important role in spectrum sharing between radar and communication systems \cite{YuanLiangJoung2021TCOMM}. RIS have been proposed to assist the design of integrated sensing and communication (ISAC) systems by providing additional degrees of freedom (DoFs) for performance improvement \cite{WangFei2021TVT,SankarDeepak2021arXiv,JiangHuang2022JSYST,LiuLiWu2022JSTSP,LiPetropulu2022ARXIV,ZhuLiLiu2023TVT,xu2023joint,yu2023active}.

In conventional RIS that essentially reflect all received energy, the phase shifts of the elements are judiciously tuned such that the signals from different paths (e.g., signals reradiated from the RIS and direct paths from the transmitter to the receiver) are combined either constructively to enhance the received signal power at the desired users, or destructively to suppress the interference to unintended users. Motivated by this capability, RIS have for the most part been proposed for applications in which maximizing spectral efficiency is the primary concern. However, the DoFs provided by an RIS can be used for other applications as well, two of which will be the focus of this paper. The first is interference mitigation for spectrum sharing. In recent years, the coexistence of radar and communication systems on shared spectrum has attracted significant attention due to the tremendous demand for additional bandwidth from the wireless sector (e.g., growing wireless connections and mobile devices) and from radar applications (e.g., remote sensing, automobile cruise control, collision avoidance). The coexistence, if improperly implemented, can cause significant interference and performance degradation for both systems \cite{NartasilpaErricolo2016RADAR,ChiriyathPaul2016TSP,WangLiGovoni2019TSP}. To mitigate the mutual interference, the use of RIS technology has been introduced into such systems \cite{WangFeiGuo2021LWC,HeCaiYu2022JSAC}. For example, one RIS is used in \cite{WangFeiGuo2021LWC} to suppress the interference from a communication system to a radar receiver by jointly optimizing the communication transmit beamformer and  RIS phase shifts to maximize the radar probability of detection, while \cite{HeCaiYu2022JSAC} employs two RISs, one at the communication transmitter and one at the receiver, to simultaneously enhance the communication signals and suppress mutual interference.

The second application we consider is the use of RIS to enhance the physical layer security (PLS) of wireless transmissions. By adjusting the phase shifts at the RIS, the signal reflected by the RIS and the direct path signal not only can be added constructively at the legitimate user (Bob), but also superimposed destructively at the Eavesdropper (Eve). As a consequence, the propagation channel between the transmitter and Bob/Eve can be enhanced/deteriorated so that the secrecy rate (SR) of the wireless transmission can be improved. The joint transmit beamformer and RIS phase shift design that maximizes the SR has been investigated in \cite{YuXuSchober2019,DongWang2020TWC,ChuHaoShi2020LWC,TangLanHan2021TVT} for the case of perfect channel state information (CSI). Scenarios with and without partial knowledge of the eavesdropper CSI were considered in \cite{BaiWangLiu2022TCOMM} and \cite{WangBaiDong2020LSP}, respectively. The authors of \cite{WangBaiDong2020LSP} proposed a two-step scheme to maximize the jamming power received at Eve while simultaneously achieving a given quality of service at Bob. A similar goal was addressed in \cite{BaiWangLiu2022TCOMM}, where the RIS location was further optimized to improve performance. An RIS is leveraged in \cite{AhmedMishra2022LSP} for secure parameter estimation at a wireless sensor network fusion center operating in the presence of an eavesdropper. In \cite{HongPanNallanathan2020TCOMM}, artificial noise (AN) is further explored in an RIS-assisted multiple-input multiple-output (MIMO) wireless communication system to enhance the system security performance. Recently, RIS have been leveraged to enhance the PLS of an integrated sensing and communication system, where a sensing target is treated as a potential eavesdropper that canintercept the information intended for the communication user \cite{SuLiuMasouros2022TWC,MishraPetropulu2022ICASSP,HuaWuSwindlehurst2022ARXIV}.

Recently, there has been growing interest in RIS that do not reflect all incoming electromagnetic energy. Such RIS include so-called Omni or Simultaneous Reflection and Transmission Reconfigurable (STAR) surfaces that not only reflect energy but also ``refract'' it to the other side of the surface for $360^\circ$ coverage \cite{XuLMD21,LiuWCMag21,ZhangTWC22,MuLGLS22,ZhangD22}, and RIS that employs active RF receivers to collect a portion of the received energy at certain elements to be used for local channel estimation or sensing purposes \cite{HuZZ21,JinZZ21,TahaAlrabeiah2021ACCESS,Zhang21SPAWC,AlexandropoulosEldar2021arXiv,AlbaneseDevoti2021arXiv,AlamzadehI22}. A hardware design for such an RIS is discussed in \cite{AlexandropoulosEldar2021arXiv} that enables the metasurface to reflect the impinging signal in a controllable manner, while simultaneously sensing a portion of it. Other types of proposed RIS apply active transmission in addition to active reception at each element, at the cost of significantly increased complexity and power consumption \cite{LongLPL21,ZhangDaiPoor2021arXiv,NguyenVLJ22,NhanNguyen22SPAWC}.

In this paper we consider RIS that controllably absorb rather than reflect all energy, although our focus is on the benefit of the absorption itself rather than using the absorbed energy for a particular purpose. There have been few hardware-oriented designs proposed for surfaces that can controllably reflect and absorb RF energy. Conceptually, adding the capability of absorption to an RIS is straightforward, for example by including a variable resistive element together with the PIN or varactor diode that controls the behavior of each RIS element. More sophisticated designs can also be applied, such as the Omni or STAR architecture, where both the electric and magnetic currents in each RIS cell are manipulated to split the signal into a reflected part and a component that is transmitted through an optically transparent medium. To achieve absorption, the ``transmitted'' signal could simply be dissipated by a terminating load rather than allowed to propagate. Similarly, the probe used to divert a portion of the incident energy in an active RIS receiver could be simply terminated rather than being demodulated and detected for sensing. Since an absorptive RIS (ARIS) does not amplify the reflected signal, no additional noise is introduced, as would be the case with an amplifier in an active RIS transmitter. The prior work on active RIS receivers such as \cite{AlexandropoulosEldar2021arXiv} is only focused on how to exploit the sensed energy, not how to exploit the reduction in reflected energy for interference mitigation, which is the focus of the work in this paper. An ARIS serves to provide a trade-off between conventional phase-only RIS and active or Omni/STAR implementations in terms of hardware complexity, power consumption, and available DoFs for shaping the propagation channel. In particular, the hardware complexity of an ARIS does not significantly exceed that required by a phase-only RIS when the absorbed energy is not employed for other purposes, and remains essentially a passive device.

The advantage of an ARIS architecture is that controlling both the phase and the attenuation of the reflected energy provides increased DoFs that are particularly well suited for applications where interference mitigation or enhancement are critical factors. While prior work has mentioned the possibility of having RIS with elements whose reflection coefficient magnitude is less than one \cite{WuZ20,ZhaoWZZ21}, this case has not been fully studied in the literature. We will investigate three scenarios where ARIS enjoy a significant advantage over conventional RIS, as described below.
\begin{enumerate}
  \item The first application explores the coexistence of both active and passive users on shared spectrum, i.e., the spectral coexistence of radar and communication systems, where the interference mitigation problem is formulated as minimizing the Frobenius norm of the interference channel matrix between a communication base station and a radar receiver (including both the direct channel and the indirect channel relayed by the ARIS). The resulting optimization problem is proved to be a convex problem and can be efficiently solved by standard numerical solvers.
  \item Unlike the first example, where the aim is to completely eliminate the cross-interference between the radar and communication system, the second example examines the spectral coexistence of multiple device-to-device (D2D) users in the absence of a base station controller. In this scenario, the RIS is employed to enhance the desired communication links while deteriorating the cross-interference among different D2D links. Specifically, a max-min criteria that maximizes the worst-case signal-to-interference-plus-noise ratio (SINR) among the D2D links is developed. This results in a non-convex constrained optimization problem, which is solved by an alternating Dinkelbach algorithm combined with semi-definite relaxation (SDR).
  \item In the third example, the use of an ARIS is explored to secure wireless transmissions in three different ways: enhance the communication link between the transmitter and the legitimate user Bob, deteriorate the channel between the transmitter and the unauthorized eavesdropper, and strengthen the link between the jammer and Eve. All three of these contribute to the system secrecy rate. We optimize the ARIS coefficients to maximize the achievable secrecy rate, which turns out to be a highly nonconvex problem which we solve by a sequential convex relaxation based Dinkelbach algorithm.
\end{enumerate}
In each case, we present numerical results that demonstrate the ability of the ARIS to mitigate undesirable interference by reducing the elements of the interference channels to be nearly zero for both spectrum sharing scenarios, a result that cannot be achieved using only adjustments of the phase of the RIS elements as with a conventional RIS implementation. In addition, our simulation results show the convergence and effectiveness of the proposed algorithms for the PLS formulation and indicate that an ARIS can play a crucial role in enhancing the security of the network. In general, our analysis and numerical results demonstrate that, when interference mitigation is a primary goal, ARIS enjoy significant benefits compared with the conventional phase-only RIS. Note that in this paper we assume knowledge of the channel state information (CSI), which is critical for optimal configuration of both conventional RIS and ARIS. Our goal is to isolate the potential benefit of absorption at the RIS compared to reflect-only RIS, eliminating potential sources of error. In practice, an accurate and low-overhead channel estimation framework will be necessary, and examples of such techniques can be found in  \cite{SwindlehurstZhouLiu2022JPROC,LiAlexandropoulos2021TCOMM,ZhangAlexandropoulos2023TCOMM}.

The remainder of the paper is organized as follows. A simple example illustrating the benefits of an ARIS is presented in Section~\ref{sec:hybridRIS}. The coexistence problem involving radar and communication systems is discussed in Section~\ref{sec:radarcomm}, and the D2D communication application is presented in Section~\ref{sec:D2D}. Section~\ref{sec:PLS} describes the application of ARIS to secure wireless transmissions, and is followed by conclusions in Section~\ref{sec:conclusion}.

\emph{Notation:} We use boldface symbols for vectors (lower case) and matrices (upper case). The symbols $(\cdot)^T$ and $(\cdot)^H$ denote the transpose and conjugate transpose operations, respectively. We use $\Vert\cdot\Vert_F$, $\Vert\cdot\Vert_2$, and $\vert\cdot\vert$ to respectively represent the Frobenius norm, vector-2 norm, and the absolute value. A complex Gaussian distribution with mean $\ubf$ and covariance matrix $\Xibf$ is denoted by $\mathcal{CN}(\ubf,\Xibf)$. The operator $\text{vec}(\Xbf)$ represents a column vector formed by stacking the columns of matrix $\Xbf$. We further let $\mathbb{C}^{M\times N}$ denote the set of $M\times N$ matrices of complex numbers, $\lambda_\text{max}(\Xbf)$ the largest eigenvalue of $\Xbf$, $\dagger$ the pseudo-inverse operator, and $\circ$ the element-wise (Hadamard) product.

\section{Absorptive Reconfigurable Intelligent Surfaces (ARIS)}
\label{sec:hybridRIS}
As mentioned above, in this paper we consider an ARIS that does not reflect (or refract) all of the energy incident on the surface. While energy absorbed by the ARIS could in principle be used for sensing or channel estimation, here we ignore such possibilities and focus on the benefits of absorption alone. For most of the RIS applications considered in the literature, where maximizing rate is the primary objective, using an RIS to absorb energy rather than increase signal power may seem counterintuitive. However, for applications where interference suppression or enhancing the secrecy of a given transmission is paramount, selective spatial absorption of the wavefield can be beneficial. As a trivial example, it may be desirable to completely absorb energy from strong interfering sources rather than further propagating their energy in the environment, or to suppress reflections of sensitive information in the direction of a potential eavesdropper. We will demonstrate however that there are many applications where only partial absorption can be beneficial, provided that the moduli of the reflection coefficients can be individually controlled.

As mentioned previously, absorption at an RIS can be achieved by a number of means, for example by adding a variable resistive element to terminate the radial waveguide stub that might otherwise be used for power detection or sampling, or placed in parallel with a varactor diode at each RIS element. As with phase control in conventional RIS, a practical implementation would likely require the reflection coefficient amplitude to be quantized, and the phase and amplitude would likely have some degree of coupling that would prevent completely independent tuning. Note that even the assumption of a unit-modulus reflection coefficient is an idealized one; the reflection coefficient amplitude can vary up to a few dB with changes in the phase, and this variation cannot be controlled in a reflect-only architecture. To make a fair comparison, we are ignoring the existence of such non-idealities in reflect-only RIS when performing our numerical studies. Thus, as in \cite{Zhang21SPAWC,AlexandropoulosEldar2021arXiv,AlbaneseDevoti2021arXiv}, our comparisons are made under ideal circumstances for both cases, and we anticipate a few dB of loss for both approaches when implemented with real hardware. We leave study of the impact of more realistic implementations for future work.

\begin{figure}[tbh!]
\centering
\includegraphics[width=3in]{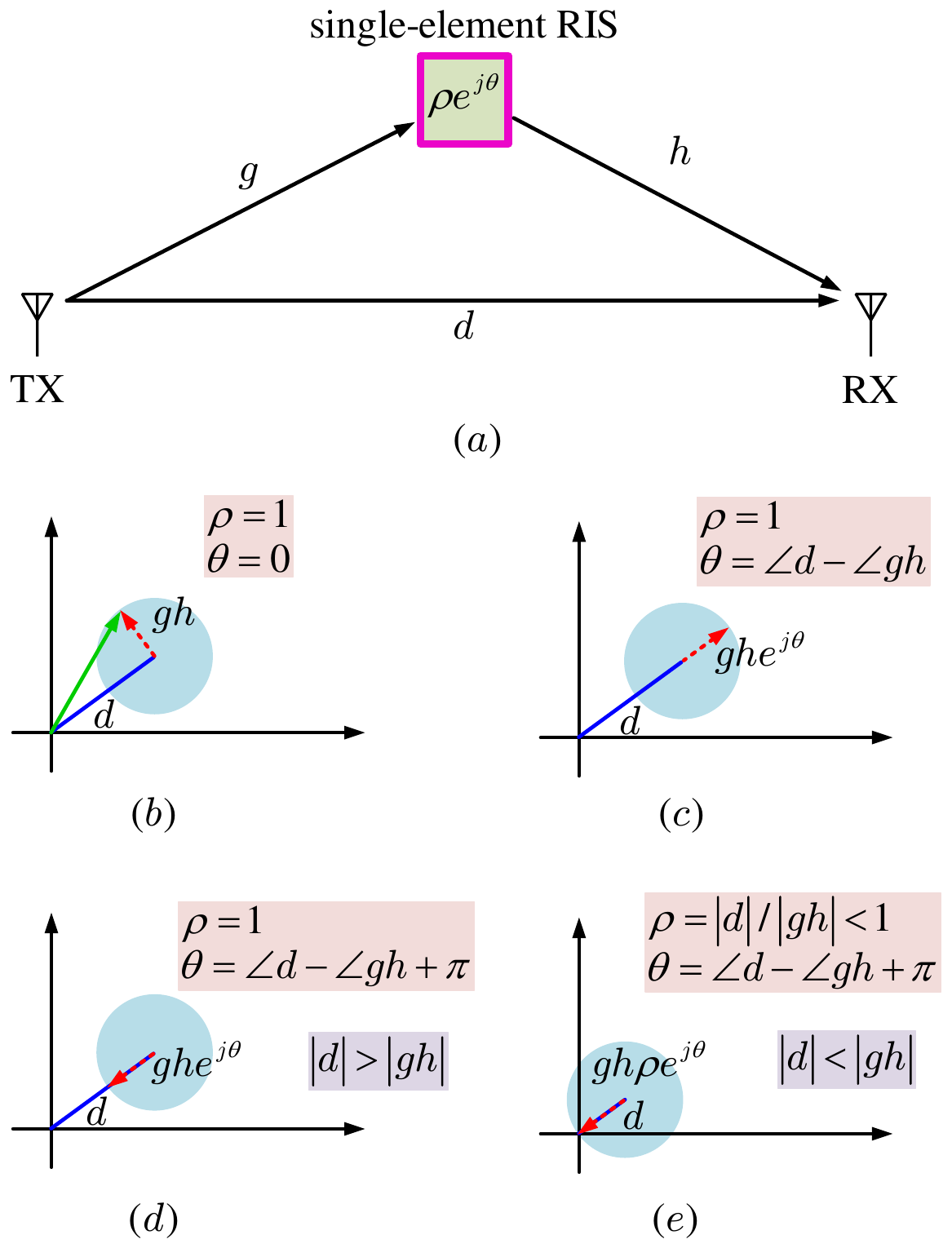}
\caption{(a) Configuration of a wireless system assisted by a single-element ARIS; (b) Combination of the direct channel and the ARIS-assisted channel when there is no ARIS design; (c) Constructive combination of the two channels with a conventional RIS design; (d) Destructive combination of the two channels with a conventional RIS design when $\vert d\vert >\vert gh\vert$; (e) Destructive combination of the two channels with an ARIS design when $\vert d\vert <\vert gh\vert$.}
\label{fgr:toyexample}
\end{figure}

To begin, we use a simple example to illustrate the benefit of an ARIS compared with a conventional phase-only RIS. Consider a single-antenna wireless system assisted by a single-element ARIS as shown in Fig.\,\ref{fgr:toyexample}(a), where $g$ denotes the channel between the transmitter (TX) and the ARIS, $h$ denotes the channel between the ARIS and the receiver (RX), and $d$ is the direct channel from the TX to the RX. The reflection coefficient of the ARIS element is represented as $\phi=\rho e^{\jmath\theta}$, with an adjustable amplitude $0 \le \rho \le 1$ and an adjustable phase, whereas in a conventional RIS it is assumed that $\rho=1$.

By judiciously tuning the ARIS reflection coefficient, the signals from different paths (e.g., signals reradiated from the RIS and the direct path from the TX to the RX) can be added either constructively to enhance the received signal power at the RX, or destructively to suppress interference at the RX. For example, Fig.~\ref{fgr:toyexample}(b) and \ref{fgr:toyexample}(c) respectively show the contributions of the direct and ARIS-reflected channels to the overall channel response between the TX and RX for two different values of $\theta$ when $\rho=1$. Clearly, to enhance the signal power at the RX, $\theta$ should be chosen as $\theta=\angle d - \angle gh$, and the conventional choice of $\rho=1$ provides the largest overall channel gain. On the other hand, if it is desired to minimize the signal power at the RX, the phase should be chosen to be in the opposite direction: $\theta=\angle d - \angle gh + \pi$ so that the direct and reflected paths add destructively. In Fig.~\ref{fgr:toyexample}(d) where $|d| > |gh|$, the conventional value $\rho=1$ provides the greatest signal suppression. However, in Fig.~\ref{fgr:toyexample}(e) where $|d| < |gh|$, which is the scenario where an ARIS is most likely to be beneficial, we see that a value of $\rho = |d|/|gh| < 1$ can completely eliminate the signal, which is only possible with an ARIS.

In the following sections we provide three more realistic scenarios where ARIS can provide significant performance gains compared with a conventional RIS. The examples are taken from problems in radar/communication system coexistence, device-to-device communications, and physical layer security.

\section{ARIS for Radar and Communication System Coexistence}
\label{sec:radarcomm}
In this section, we consider the interference suppression problem for the coexistence of radar and communication systems on shared spectrum with a deployed ARIS. We study the benefit of employing an ARIS to mitigate the mutual interference between the co-channel radar and communication systems. In the following, we first introduce the system model and formulate the interference mitigation problem for spectrum sharing between the radar and communication systems. Then, the resulting non-linear constrained optimization problem is proved to be a convex problem, and the formulation and solution for the problem using a conventional RIS are provided. Finally, numerical results are provided to verify the effectiveness of the ARIS in suppressing the mutual interference between the two systems.

\subsection{Problem Formulation}
\begin{figure}[tbh!]
\centering
\includegraphics[width=3.3in]{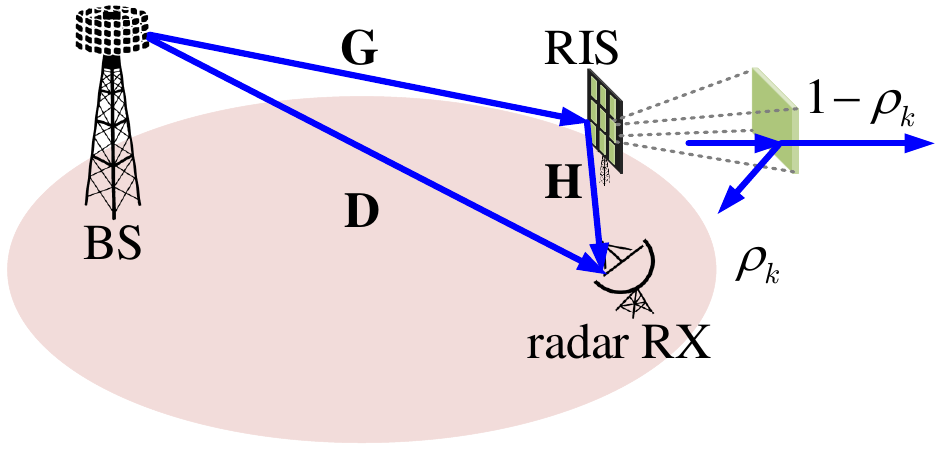}
\caption{RIS-Assisted interference suppression for the coexistence of radar and communication systems on shared spectrum.}
\label{fgr:radarcommconfiguration}
\end{figure}
Consider the coexistence of a communication base station (BS) with $M$ antennas and a radar RX with $N$ antennas using the same frequency band at the same time and/or at the same location. An ARIS with $K$ reflecting elements is deployed near the radar RX to mitigate the interference from the BS to the radar RX as shown in Fig.\,\ref{fgr:radarcommconfiguration}. As introduced in Section~\ref{sec:hybridRIS}, the ARIS employs elements with an adjustable modulus coefficient $\rho_k\in[0\,1]$. Let $\theta_k$ denote the phase shift associated with the $k$-th RIS element, so that the diagonal matrix accounting for the response of the ARIS can be expressed as $\Phibf\triangleq\text{diag}(\rho_1e^{\jmath\theta_1},\cdots,\rho_Ke^{\jmath \theta_K})$. The channel between the BS and radar receiver can be written as $\Dbf+\Hbf\Phibf\Gbf$, where $\Dbf\in\mathbb{C}^{N\times M}$, $\Gbf\in\mathbb{C}^{K\times M}$, and $\Hbf\in\mathbb{C}^{N\times K}$ respectively denote the channel matrices between the BS and radar RX, the BS and the ARIS, and the ARIS and the radar RX.

The problem of interest is to suppress the interference channel from the BS to the radar RX by jointly designing the phase shifts $\theta_k$ and moduli $\rho_k$ of the ARIS for $k=1,\cdots,K$. Specifically, the interference suppression problem can be formulated as
\begin{subequations}\label{eq:radarcomm}
\begin{gather}
\min_{\{\rho_k,\theta_k\}}\quad\Big\Vert\Dbf+\Hbf\Phibf\Gbf\Big\Vert_F\label{eq:radarcomm_cost}\\
\text{s.t.}~\Phibf=\text{diag}(\rho_1e^{\jmath\theta_1},\cdots,\rho_Ke^{\jmath \theta_K}),\label{eq:radarcomm_cons_1}\\
0\leq\rho_k\leq1,~\forall k.\label{eq:radarcomm_cons_2}
\end{gather}
\end{subequations}
Note that there are many metrics that have been proposed to evaluate radar performance in integrated sensing and communication (ISAC) systems, including the Cram\'er-Rao Bound (CRB) \cite{LiuMasouros2022TSP,song2022cramerrao} and detection probability \cite{AnLiChau2023TWC}. In the approach presented here, the norm of the cross interference channel from the BS to the radar RX is minimized to improve the received radar SINR, which in turn improves other radar metrics such as the probability of detection. In the following, we will demonstrate that the extra DoFs provided by $\rho_k$ enables the interference channel in principle to be completely eliminated (i.e., $\Dbf+\Hbf\Phibf\Gbf \rightarrow 0$) with a sufficiently large number of ARIS elements when the ARIS-assisted signal path is stronger than that of the direct path $\Dbf$. This is not possible by only adjusting the phase of the RIS elements as in a conventional RIS.

\subsection{Proposed Solutions}
In this section, we solve the above optimization problem for both the ARIS and conventional RIS.

\subsubsection{ARIS Design}
While the interference mitigation problem in \eqref{eq:radarcomm} is a nonlinear constrained optimization problem whose solution cannot be obtained directly, it can be reformulated to be convex and thus a global optimum can be found using standard techniques. In the following, we simplify the problem via mathematical manipulation. Specifically, by defining $\phibf\triangleq[\rho_1e^{\jmath\theta_1},\cdots,\rho_Ke^{\jmath \theta_K}]^T$, the two constraints in \eqref{eq:radarcomm_cons_1} and \eqref{eq:radarcomm_cons_2} can be combined as the following convex constraint:
\ben\label{eq:constriant}
\vert\phibf(k)\vert\leq 1,~\forall k,
\een
where $\phibf(k)$ is the $k$-th element of $\phibf$, i.e., $\phibf(k)=\rho_ke^{\jmath \theta_k}$. Then, the optimization problem in \eqref{eq:radarcomm} is equivalent to
\begin{subequations}\label{eq:radarcomm_P2}
\begin{gather}
\min_{\phibf}~~\Big\Vert\Dbf+\Hbf\text{diag}(\phibf)\Gbf\Big\Vert_F\label{eq:radarcommP2_cost}\\
\text{s.t.}~\vert\phibf(k)\vert\leq 1,~\forall k.
\end{gather}
\end{subequations}
In addition, we have the following transformation for the cascaded channel:
\ben
\Hbf\text{diag}(\phibf)\Gbf=\sum_{k=1}^K\phibf(k)\Hbf(:,k)\Gbf(k,:),
\een
where $\Hbf(:,k)$ is the $k$-th column of $\Hbf$ and $\Gbf(k,:)$ is the $k$-th row of $\Gbf$.

The cost function in \eqref{eq:radarcommP2_cost} can be rewritten as
\begin{align}
&\Big\Vert\Dbf+\sum_{k=1}^K\phibf(k)\Hbf(:,k)\Gbf(k,:)\Big\Vert_F\stackrel{(a)}{=}\Big\Vert\Dbf+\sum_{k=1}^K\phibf(k)\Ybf_k\Big\Vert_F\notag\\
&\stackrel{(b)}{=}\Big\Vert\text{vec}(\Dbf)+\sum_{k=1}^K\phibf(k)\text{vec}(\Ybf_k)\Big\Vert_2\stackrel{(c)}{=}\Big\Vert\dbf+\Abf\phibf\Big\Vert_2,
\end{align}
where $\dbf=\text{vec}(\Dbf)$. Note that the equality in (a) is obtained by letting $\Ybf_k=\Hbf(:,k)\Gbf(k,:)$, the equality in (b) is achieved by transforming the matrices into their vector forms, and the equality in (c) results from defining $\Abf$ as the $MN\times K$ matrix whose $k$-th column is given by $\Abf(:,k)=\text{vec}(\Ybf_k)$. With these definitions, the original optimization problem in \eqref{eq:radarcomm} can be reformulated as
\begin{subequations}\label{eq:radarcomm_P3}
\begin{gather}
\min_{\phibf}~~\Big\Vert\dbf+\Abf\phibf\Big\Vert_2\\
\text{s.t.}~\vert\phibf(k)\vert\leq 1,~\forall k,
\end{gather}
\end{subequations}
which is a convex problem that can be efficiently solved.

\subsubsection{Design with Conventional RIS}
The problem formulation using the ARIS can be easily transformed to that for a conventional RIS by changing the constraint in \eqref{eq:radarcomm_P3} to $\vert\phibf(k)\vert=1,\forall k$, i.e.,
\begin{subequations}\label{eq:radarcomm_P4}
\begin{gather}
\min_{\phibf}~~\Big\Vert\dbf+\Abf\phibf\Big\Vert_2\\
\text{s.t.}~\vert\phibf(k)\vert=1,~\forall k,
\end{gather}
\end{subequations}
which is a nonconvex problem due to the constant-modulus constraint. It can be recast as unit-modulus constrained quadratic program and solved by applying the semidefinite relaxation (SDR) technique \cite{LuoMa2010MSP}. Recently, a fast solution to least squares problems with unit-modulus constraints was proposed in \cite{TranterSidiropoulo2017} based on the gradient projection scheme. For completeness, we summarize the process of solving problem \eqref{eq:radarcomm_P4} in Algorithm \ref{alg:GP} below.
\begin{algorithm}
\caption{Gradient Projection Method to Solve \eqref{eq:radarcomm_P4}}
\begin{algorithmic}
\label{alg:GP}
\STATE \textbf{input:} Channel matrices $\Hbf$, $\Gbf$, and $\Dbf$.
\STATE  \textbf{initialization:} Set the iteration index $i=0$, $\phibf^{(0)}=e^{\jmath\angle(-\Abf^\dagger\dbf)}$, and $\beta=\frac{0.9}{\lambda_\text{max}(\Abf^H\Abf)}$.
\\
\REPEAT
\STATE
\begin{enumerate}
\item Gradient: $\xibf^{(i+1)}=\phibf^{(i)}-\beta\Abf^H(\dbf+\Abf\phibf^{(i)})$.
\item Projection: $\phibf^{(i+1)}=e^{\jmath\angle(\xibf^{(i+1)})}$.
\item Set $i=i+1$
\end{enumerate}
\UNTIL convergence.
\end{algorithmic}
\end{algorithm}

\subsection{Numerical Results}
In this section, simulation results are presented to evaluate the capability of the ARIS in interference mitigation. The performance of the ARIS (denoted as \textbf{ARIS}) is compared with that of a conventional  RIS (denoted as \textbf{conv. RIS}), and each result is based on 2500 random channel realizations unless otherwise stated.

\begin{figure}
\centering
\includegraphics[width=3.3in]{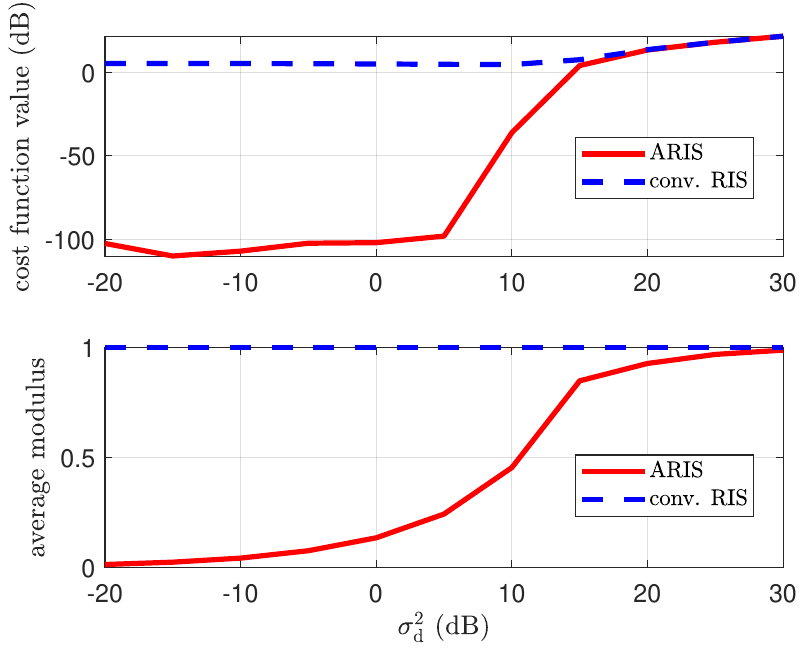}
\caption{Performance of ARIS in interference suppression for the coexistence of radar and communication systems for various direct channel strengths $\sigma^2_\text{d}$ when $\sigma^2_\text{h}=\sigma^2_\text{g}=0$ dB and $K=64$. Top: Norm of interference channel versus $\sigma^2_\text{d}$; Bottom: Average ARIS modulus value ($\rho_k$) versus $\sigma^2_\text{d}$.}
\label{fgr:results_radarcomm_sigma_D}
\end{figure}
First, we consider the coexistence of a communication BS with $M=6$ antennas and a radar receiver with $N=6$ antennas, where independent identically distributed ({i.i.d.}) Rayleigh fading channel models are assumed for $\Hbf$, $\Gbf$, and $\Dbf$ with channel variances $\sigma_\text{h}^2$, $\sigma_\text{g}^2$, and $\sigma_\text{d}^2$, respectively. Fig.\,\ref{fgr:results_radarcomm_sigma_D} shows the performance of the proposed schemes versus various direct-path channel strengths between the communication transmitter and radar receiver $\sigma_\text{d}^2$ when $\sigma_\text{h}^2=\sigma_\text{g}^2=0$ dB and the number of RIS elements is $K=64$. It can be seen from the top subfigure of Fig.\,\ref{fgr:results_radarcomm_sigma_D} that the ARIS is able to completely suppress the interference channel, forcing the norm of the channel $\Vert\Dbf+\Hbf\boldsymbol{\Phi}\Gbf\Vert_F$ to essentially zero when $\sigma_\text{d}^2\leq10$ dB.  We see that this result can not be achieved using the conventional RIS, which leads to an interference channel with non-negligible strength. This is because when the direct-path channel strength is smaller than that of the RIS-relayed channel, we have a situation similar to Fig.~\ref{fgr:toyexample}(e), and the ARIS has a sufficient number of DoFs to force $\Hbf\boldsymbol{\Phi}\Gbf+\Dbf \rightarrow 0$. This is also verified by the bottom subfigure of Fig.\,\ref{fgr:results_radarcomm_sigma_D}, where the average optimal modulus of the ARIS elements is plotted. When $\sigma_\text{d}^2$ is very small, the ARIS absorbs most of the incident energy, since very little is required to cancel the interference at the radar RX. In the limit where $\sigma_\text{d}^2 \rightarrow 0$, the obvious solution is simply to ``turn off'' the ARIS. As $\sigma_\text{d}^2$ increases, the interference channel strength for both the ARIS and conventional RIS increases, but there is a large range of $\sigma_\text{d}^2$ for which the ARIS leads to significantly reduced interference. Eventually, the performance of the ARIS and conventional RIS become identical for large $\sigma_\text{d}^2$ since the full strength of the ARIS-relayed channel is needed to cancel the direct-path channel as much as possible, similar to what was seen in Fig.~\ref{fgr:toyexample}(d). This is also shown in the bottom subfigure of Fig.\,\ref{fgr:results_radarcomm_sigma_D} where the average modulus of the ARIS gets closer to 1 as $\sigma_\text{d}^2$ increases and eventually becomes 1 at $\sigma_\text{d}^2=30$ dB.

\begin{figure}
\centering
\includegraphics[width=3.3in]{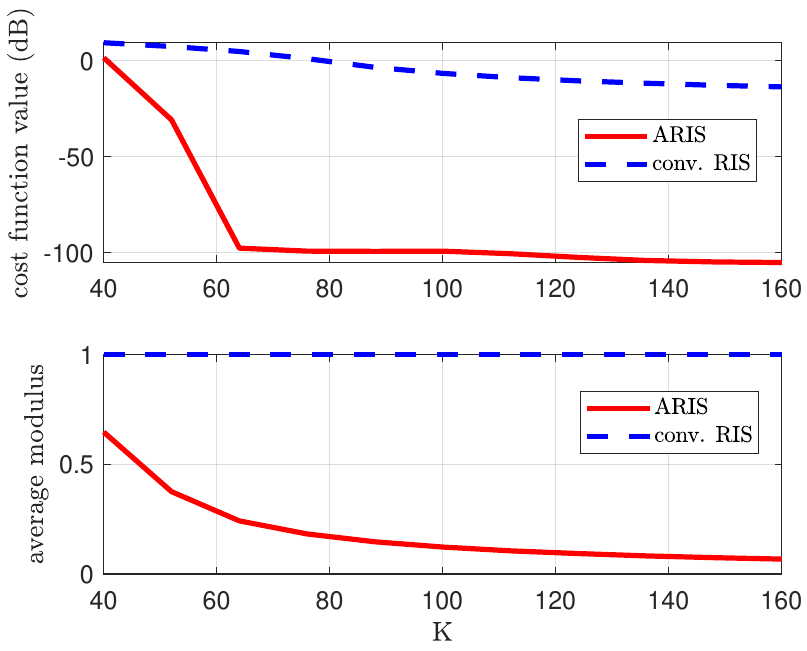}
\caption{Performance of ARIS in interference suppression for the coexistence of radar and communication systems versus the number of RIS elements $K$ when $\sigma^2_\text{h}=\sigma^2_\text{g}=0$ dB and $\sigma^2_\text{d}=5$ dB. Top: Norm of interference channel versus $\sigma^2_\text{d}$; Bottom: Average RIS modulus value ($\rho_k$) versus $\sigma^2_\text{d}$.}
\label{fgr:results_radarcomm_K}
\end{figure}

Fig.\,\ref{fgr:results_radarcomm_K} shows the interference suppression performance versus the number of ARIS elements $K$ when $\sigma^2_\text{h}=\sigma^2_\text{g}=0$ dB and $\sigma^2_\text{d}=5$ dB. The ARIS outperforms the conventional RIS for all $K$, especially for larger ARIS due to the availability of extra DoFs from the adaptivity of the RIS element modulus which enables the ARIS to completely suppress the interference. Increasing $K$ also improves the performance of the ability of the conventional RIS to reduce the interference, but compared with the ARIS a significant residual remains.

The results above were obtained for full-rank Rayleigh fading channels. In the next example, we examine the performance of the proposed ARIS in interference suppression issue under a millimeter wave (mmW) propagation model in which each channel is composed of $T$ angular clusters of closely spaced signal arrivals. In particular, the channels $\Dbf$, $\Hbf$, and $\Gbf$ are generated as \cite{AkdenizLiu2014JSAC}:
\begin{align}
\Dbf&=\sqrt{NM}\sum_{t=1}^T\sum_{\ell=1}^J\alpha_{\text{d},t,\ell}\abf_\text{rx}(\varphi_{t,\ell}^\text{rx,tx})\abf_\text{tx}(\widetilde\varphi_{t,\ell}^\text{rx,tx})^T,\notag\\
\Gbf&=\sqrt{KM}\sum_{t=1}^T\sum_{\ell=1}^J\alpha_{\text{g},t\ell}\abf_\text{ris}(\varphi_{t,\ell}^\text{ris,tx})\abf_\text{tx}(\widetilde\varphi_{t,\ell}^\text{ris,tx})^T,\notag\\
\Hbf&=\sqrt{NK}\sum_{t=1}^T\sum_{\ell=1}^J\alpha_{\text{h},t,\ell}\abf_\text{rx}(\varphi_{t,\ell}^\text{rx,ris})\abf_\text{ris}(\widetilde\varphi_{t,\ell}^\text{rx,ris})^T,
\end{align}
where $J$ is the number subpath within each cluster, $\alpha_{\text{d},t,\ell}\sim\mathcal{CN}(0,\sigma_\text{d}^2)$, $\alpha_{\text{g},t,\ell}\sim\mathcal{CN}(0,\sigma_\text{g}^2)$, and $\alpha_{\text{h},t,\ell}\sim\mathcal{CN}(0,\sigma_\text{h}^2)$ are random complex channel gains, $\varphi_{t,\ell}^\text{p,q}$ denotes the angle of arrival (AoA) for a signal arriving at $\text{p}\in\{\text{RX,RIS}\}$ that originated from $\text{q}\in\{\text{TX,RIS}\}$, $\widetilde\varphi_{t,\ell}^\text{p,q}$ is similarly defined for the angle of departure (AoD), and $\abf_\text{tx}(\cdot)\in\mathbb{C}^{M\times1}$, $\abf_\text{ris}(\cdot)\in\mathbb{C}^{K\times1}$, and $\abf_\text{rx}(\cdot)\in\mathbb{C}^{N\times1}$ are the normalized steering vectors for the TX, RIS, and RX antenna arrays, respectively. For this simulation, we assume a uniform linear array with half-wavelength elements spacing for the TX, RIS, and RX, in which the steering vector is expressed as
\ben
\abf_\text{tx}(\varphi)=\frac{1}{\sqrt{M}}[1,e^{\jmath \pi\sin\varphi},\cdots,e^{\jmath (M-1)\pi\sin\varphi}]^T ,
\een
and similarly for the other arrays.
In the simulation, all the AoAs and AoDs are randomly generated. In particular, we take the generation of $\varphi_{t,\ell}^{\text{rx,tx}}$ as an example: the central angle of the $\ell$-th cluster for the AoA between the RX and TX (denoted as $\varphi_{\ell}^{\text{c,rx,tx}}$) is uniformly distributed within $60^\circ$ of a predefined angle (denoted as $\vartheta^{\text{rx,tx}}$)\footnote{$\vartheta^{\text{p,q}}$ denotes the predefined angle for the AoA at $\text{p}\in\{\text{RX},\text{RIS}\}$ that originated from $\text{q}\in\{\text{TX},\text{RIS}\}$, $\widetilde\vartheta^{\text{p,q}}$ is similarly defined for the predefined angle of AoD. In the simulation, the predefined angles are chosen based on the relative locations of the TX, RIS, and RX.}, i.e., $\varphi_{\ell}^{\text{c,rx,tx}}\in[\varphi^{\text{rx,tx}}-60^\circ,\varphi^{\text{rx,tx}}+60^\circ]$. Then, the subpaths in this cluster are randomly generated within $2^\circ$ of the cluster's central angle, i.e., $\varphi_{t,\ell}^{\text{rx,tx}}\in[\varphi_{\ell}^{\text{c,rx,tx}}-2^\circ,\varphi_{\ell}^{\text{c,rx,tx}}+2^\circ]$. In the simulation, we have $\widetilde\vartheta^\text{rx,tx}=15^\circ$, $\widetilde\vartheta^\text{ris,tx}=30^\circ$, $\widetilde\vartheta^\text{rx,ris}=-15^\circ$, $\vartheta^\text{rx,tx}=\widetilde\vartheta^\text{rx,tx}+180^\circ$, $\vartheta^\text{ris,tx}=\widetilde\vartheta^\text{ris,tx}+180^\circ$, and $\vartheta^\text{rx,ris}=\widetilde\vartheta^\text{rx,ris}+180^\circ$. In general, the larger the number of clusters the higher the rank of the channel matrix.

\begin{figure}
\centering
\includegraphics[width=3.3in]{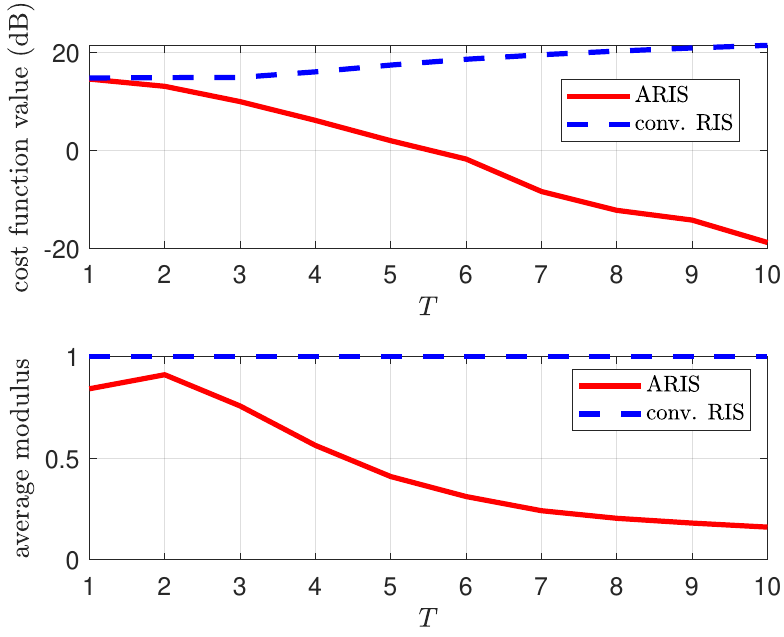}
\caption{Performance of ARIS in interference suppression for the coexistence of radar and communication systems using a mmW channel model for various numbers of channel clusters when $\sigma^2_\text{h}=\sigma^2_\text{g}=0$ dB, $\sigma^2_\text{d}=10$ dB, $K=64$, and $J=4$. Top: Norm of interference channel versus number of clusters $T$; Bottom: Average RIS modulus value ($\rho_k$) versus $T$.}
\label{fgr:results_radarcomm_T}
\end{figure}

The performance of the two types of RIS is plotted in Fig.~\ref{fgr:results_radarcomm_T} against the number of the channel clusters $T$ when $\sigma^2_\text{h}=\sigma^2_\text{g}=0$ dB, $\sigma^2_\text{d}=10$ dB, $K=64$, and $J=4$. We see that when only one or two clusters are present, the ARIS is not able to achieve a significant benefit compared to the standard RIS as the subspaces of the channels are likely not aligned, and thus cannot be combined destructively. Note that the average modulus of the ARIS is slightly lower for $T=1$ than for $T=2$. This can be explained as follows. When $T=1$ and $J=4$, both the $6 \times 6$ direct channel $\mathbf{D}$ and the $6 \times 6$ ARIS channel $\mathbf{H}\boldsymbol{\Phi}\mathbf{G}$ will be at most rank 4, and the subspaces spanned by these channels will in general be different. This further means that the null space of the direct channel is not orthogonal to the ARIS channel subspace, and thus some of the energy from the ARIS channel will actually produce interference in this null space that would not have existed if the ARIS were not present. Thus, the ARIS must balance its interference cancellation benefit in the overlapping signal subspace with the extra interference it produces in the orthogonal null space, and hence the optimal solution in this case is to actually absorb more energy than when $T=2$ in favor of producing slightly less null space interference. When $T$ increases to 2, the channels are now generically full rank and there is no null space remaining in which the RIS channel will cause extra interference. The benefit of absorption is more apparent for $T\ge 3$ since the strength of the channel subspaces becomes more uniformly distributed, and the RIS has more options available to configure itself for interference cancellation. As the number of clusters increases further, the results mirror those previously given for Rayleigh fading channels.

\section{ARIS for Device-to-Device Communications}
\label{sec:D2D}
In this section, we investigate spectrum sharing for D2D communications where the ARIS is deployed to suppress the mutual interference among D2D links. We first formulate the D2D ARIS-assisted interference mitigation scenario as a max-min problem that considers the worst-case SINR among the $L$ links as the figure of merit to be optimized. Then, we develop the solution for the resulting optimization problem, which is followed by a discussions of how to employ conventional RIS for the design.

\subsection{Problem Formulation}
\begin{figure}[tbh!]
\centering
\includegraphics[width=3.3in]{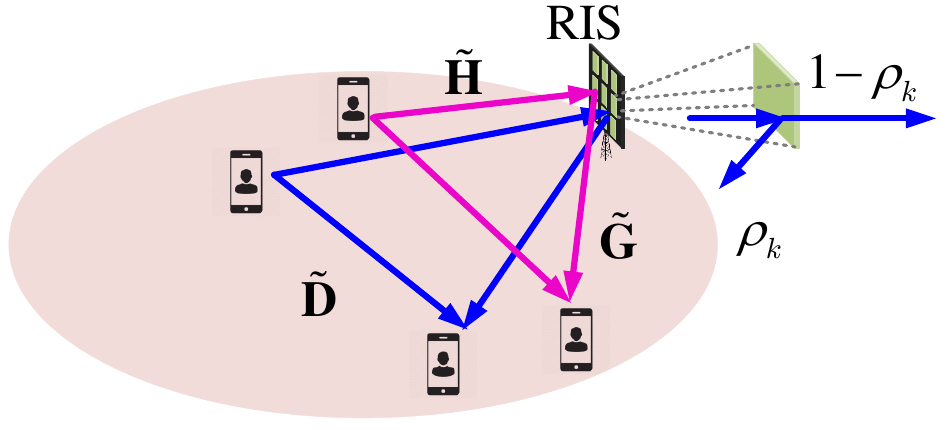}
\caption{RIS-Assisted cross-interference mitigation for spectrum sharing in device-to-device communication.}
\label{fgr:D2Dconfiguration}
\end{figure}
Device-to-device communications can increase network spectral efficiency, reduce transmission delay, and alleviate traffic congestion for cellular infrastructure by enabling users in close proximity to directly transmit signals to each other without using a BS as a relay. In this section, we consider the cross-interference management problem among multiple single-antenna D2D communication links over shared spectrum with a deployed ARIS as shown in Fig.\,\ref{fgr:D2Dconfiguration}, where $L$ D2D pairs coexist with each other in the same frequency band. Similar to the coexistence of radar and communication systems in Section~\ref{sec:radarcomm}, an ARIS with a reflection coefficient matrix $\Phibf$ is deployed to mitigate the cross interference among the D2D links.

Let $\widetilde{\Dbf}\in\mathbb{C}^{L\times L}$, $\widetilde{\Gbf}\in\mathbb{C}^{K\times L}$, and $\widetilde{\Hbf}\in\mathbb{C}^{L\times K}$ respectively denote the channel matrices between the TXs and RXs, the TXs and the RIS, and the RIS and the RXs, so that the overall channel between the TXs and RXs of the D2D links can be expressed as $\widetilde{\Hbf}\Phibf\widetilde{\Gbf}+\widetilde{\Dbf}$. Unlike the case in Section \ref{sec:radarcomm} where the goal was to completely eliminate the interference channel $\Hbf\Phibf\Gbf+\Dbf$, here we need to eliminate (or at least substantially reduce) the off-diagonal elements of the channel matrix $\widetilde{\Hbf}\Phibf\widetilde{\Gbf}+\widetilde{\Dbf}$ since they correspond to the mutual interference among the D2D links. To guarantee the performance of all D2D communication links, the cross-interference mitigation problem is formulated as maximizing the minimum SINR of the $L$ links by designing the parameters of the ARIS, i.e.,
\begin{subequations}\label{eq:D2D}
\begin{gather}
\max_{\{\rho_k,\theta_k\}}~\min_{\ell=1,\cdots,L}~\frac{f_{\ell,\ell}(\rho_k,\theta_k)}{\sum_{\bar{\ell}=1,\bar{\ell}\neq \ell}^Lf_{\ell,\bar{\ell}}(\rho_k,\theta_k)+\sigma^2}\\
\text{s.t.}~\eqref{eq:radarcomm_cons_1},~ \eqref{eq:radarcomm_cons_2},\label{eq:D2D_const}
\end{gather}
\end{subequations}
where the noise variance $\sigma^2$ is assumed to the be the same for all RXs,
\ben
f_{\ell,\bar{\ell}}(\rho_k,\theta_k)=\big\vert\widetilde{\Hbf}(\ell,:)\Phibf\widetilde{\Gbf}(:,\bar{\ell})+\widetilde{\Dbf}(\ell,\bar{\ell})\big\vert^2p_{\bar\ell}^2 \; ,
\een
$p_{\bar\ell}^2$ is the transmit power of the $\bar\ell$-th link, and  $\widetilde{\Dbf}(\ell,\bar{\ell})$ is the $(\ell,\bar\ell)$-th element of $\widetilde{\Dbf}$. The above problem is nonconvex and cannot be solved in closed-form. In the following, an iterative method is employed to solve it.

\subsection{Proposed Solution}
To solve \eqref{eq:D2D}, the definition of $\phibf(k)=\rho_ke^{\jmath\theta_k}$ in Section \ref{sec:radarcomm} is employed to simplify the problem, in which case the constraint \eqref{eq:D2D_const} can be replaced by \eqref{eq:constriant} and $f_{\ell,\bar{\ell}}(\rho_k,\theta_k)$ can be equivalently expressed as
\begin{align}\label{eq:f}
f_{\ell,\bar{\ell}}(\rho_k,\theta_k)&=\big\vert\widetilde\Hbf(\ell,:)\Phibf\widetilde\Gbf(:,\bar{\ell})+\widetilde\Dbf(\ell,\bar{\ell})\big\vert^2p_{\bar\ell}^2\notag\\
&=\big\vert\hbf_{\ell,\bar{\ell}}^H\phibf+\widetilde\Dbf(\ell,\bar{\ell})\big\vert^2p_{\bar\ell}^2\notag\\
&=\bar{\phibf}^H\Fbf_{\ell,\bar{\ell}}\bar{\phibf}p_{\bar\ell}^2,
\end{align}
where $\hbf_{\ell,\bar{\ell}}^{\ast}(k)=\widetilde\Hbf(\ell,k)\widetilde\Gbf(k,\bar{\ell})$, $\bar{\phibf}=[\phibf^T~1]^T$, and
\begin{equation}
\begin{split}
\Fbf_{\ell,\bar{\ell}}=
\begin{bmatrix}
\hbf_{\ell,\bar{\ell}}\hbf_{\ell,\bar{\ell}}^H& \hbf_{\ell,\bar{\ell}} \widetilde\Dbf_{\ell,\bar{\ell}}\\
\hbf_{\ell,\bar{\ell}}^H \widetilde\Dbf_{\ell,\bar{\ell}}^\ast& \vert \widetilde\Dbf_{\ell,\bar{\ell}}\vert^2
\end{bmatrix}.
\end{split}
\end{equation}
Note that the third equality in \eqref{eq:f} is achieved by introducing the auxiliary scalar $1$ in $\bar{\phibf}$ to convert the non-homogeneous quadratic program into a homogeneous one.

The optimization problem in \eqref{eq:D2D} can be rewritten as
\begin{subequations}\label{eq:D2D_P1}
\begin{gather}
\max_{\bar\phibf}~\min_{\ell=1,\cdots,L}~\frac{\bar{\phibf}^H\Fbf_{\ell,\ell}\bar{\phibf}p_{\ell}^2}{\sum_{\bar{\ell}=1,\bar{\ell}\neq \ell}^L\bar{\phibf}^H\Fbf_{\ell,\bar{\ell}}\bar{\phibf}p_{\bar\ell}^2+\sigma^2}\\
\text{s.t.}~\vert\bar\phibf(k)\vert\leq 1, k=1,\cdots,K,~\text{and}~\vert\bar\phibf(K+1)\vert= 1\label{eq:D2D_P1_const}.
\end{gather}
\end{subequations}
This is a fractional quadratically constrained quadratic programming (QCQP) problem and can be solved using SDR along with the Dinkelbach algorithm. In the following, we first use SDR to convert \eqref{eq:D2D_P1} to a fractional programming problem by dropping the rank-one constraint. Let $\bar\Phibf=\bar\phibf\bar\phibf^H$ to obtain the SDR form of \eqref{eq:D2D_P1} as
\begin{subequations}\label{eq:D2D_SDR}
\begin{gather}
\max_{\bar\Phibf}\,\min_{\ell=1,\cdots,L}\,\frac{\text{tr}(\bar{\Phibf}\Fbf_{\ell,\ell})p_{\ell}^2}{\sum_{\bar{\ell}=1,\bar{\ell}\neq \ell}^L\text{tr}(\bar{\Phibf}\Fbf_{\ell,\bar{\ell}})p_{\bar\ell}^2+\sigma^2}\\
\text{s.t.}\,\vert\bar\Phibf(k,k)\vert\leq 1, k=1,\cdots,K,\label{eq:D2D_SDR_const1}\\
\vert\bar\Phibf(K+1,K+1)\vert= 1,\label{eq:D2D_SDR_const2}
\end{gather}
\end{subequations}
which is a fractional programming problem and can be solved by the Dinkelbach algorithm in polynomial time \cite{Dinkelbach67}. Specifically, by introducing a slack variable $\lambda$, problem \eqref{eq:D2D_SDR} becomes
\begin{subequations}\label{eq:D2D_Dinkelbach}
\begin{gather}
\max_{\bar\Phibf,\lambda}\,\min_{\ell=1,\cdots,L}\,\text{tr}(\bar{\Phibf}\Fbf_{\ell,\ell})p_{\ell}^2-\lambda\left(\sum_{\bar{\ell}=1,\bar{\ell}\neq \ell}^L\text{tr}(\bar{\Phibf}\Fbf_{\ell,\bar{\ell}})p_{\bar\ell}^2+\sigma^2\right)\\
\text{s.t.}\,\eqref{eq:D2D_SDR_const1},\,\eqref{eq:D2D_SDR_const2}.
\end{gather}
\end{subequations}

The above problem can be solved by alternatingly optimizing the cost function with respect to (w.r.t.) $\bar\Phibf$ and $\lambda$. In particular, by fixing $\lambda$ to the value obtained from the $i$-th iteration, $\lambda^{(i)}$, we can obtain $\bar\Phibf^{(i+1)}$ by solving
\begin{subequations}\label{eq:D2D_Dinkelbach_Theta}
\begin{gather}
\max_{\bar\Phibf}\,\min_{\ell=1,\cdots,L}\,\text{tr}(\bar{\Phibf}\Fbf_{\ell,\ell})p_{\ell}^2-\lambda^{(i)}\left(\sum_{\bar{\ell}=1,\bar{\ell}\neq \ell}^L\text{tr}(\bar{\Phibf}\Fbf_{\ell,\bar{\ell}})p_{\bar\ell}^2+\sigma^2\right)\\
\text{s.t.}\,\eqref{eq:D2D_SDR_const1},\,\eqref{eq:D2D_SDR_const2},
\end{gather}
\end{subequations}
which is a convex problem. Next, we find $\lambda$ by fixing $\bar\Phibf$ to the value obtained from the latest update, $\bar\Phibf^{(i+1)}$, in which case $\lambda^{(i+1)}$ is solved in closed-form as
\ben
\lambda^{(i+1)}=\arg\,\min_{\ell,=1,\cdots,L}\,\frac{\text{tr}(\bar{\Phibf}\Fbf_{\ell,\ell})p_{\ell}^2}{\sum_{\bar{\ell}=1,\bar{\ell}\neq \ell}^L\text{tr}(\bar{\Phibf}\Fbf_{\ell,\bar{\ell}})p_{\bar\ell}^2+\sigma^2}.
\een
The alternating process is repeated until the algorithm converges, i.e., the improvement of the cost function over two successive iterations is smaller than a predefined tolerance $\epsilon$. After convergence, the optimum solution is denoted as $\hat\Phibf$.

After solving \eqref{eq:D2D_SDR} through the iterative Dinkelbach algorithm, we need to convert the optimum solution $\hat\Phibf$ to \eqref{eq:D2D_SDR} into a feasible solution $\hat\phibf$ to \eqref{eq:D2D_P1}. This can be achieved through a randomization approach \cite{LuoMa2010MSP}. Specifically, given the optimum $\hat\Phibf$, a set of Gaussian random vectors are generated, i.e., $\xibf_j\sim\mathcal{CN}(\mathbf{0},\hat\Phibf)$, $j=1,\cdots,J$, where $J$ is the number of randomization trials. Since the $\xibf_j$ are not always feasible for the modulus constraints in \eqref{eq:D2D_P1}, we need to first normalize them as $\bar\xibf_j=\xibf_j/\xibf_j(K+1)$ to satisfy the constraint $\vert\bar\phibf(K+1)\vert=1$. Then, to meet the constraint $\vert\bar\phibf(k)\vert\leq1$, the feasible solution can be further recovered by $\hat\xibf_j(k)=e^{\jmath\angle(\bar\xibf_j(k))}$ if $\vert\bar\xibf_j(k)\vert>1$. Finally, the rank-one solution can be obtained as
\ben
\hat\phibf=\arg\max_{\{\hat\xibf_j\}}~\min_{\ell=1,\cdots,L}~\frac{\hat\xibf_j^H\Fbf_{\ell,\ell}\hat\xibf_jp_{\ell}^2}{\sum_{\bar{\ell}=1,\bar{\ell}\neq \ell}^L\hat\xibf_j^H\Fbf_{\ell,\bar{\ell}}\hat\xibf_jp_{\bar\ell}^2+\sigma^2}.
\een

Similar to the design for the coexistence of radar and communication systems in Section \ref{sec:radarcomm}, the formulation for the design using the standard RIS can be easily obtained by changing the constraint \eqref{eq:D2D_P1_const} to $\vert\bar\phibf(k)\vert=1$ for $k=1,\cdots, K+1$. The solution can be accordingly adjusted with the new constraint without imposing additional difficulties.

The computational complexity of the proposed algorithm depends on the number of iterations $L_1$ of the Dinkelbach algorithm, and the number of randomized trials $J$ for the semidefinite relaxation. Simulation results show that the number of iterations required by the algorithm is relatively small, e.g., about 10 iterations. In addition, a convex problem is solved for each iteration with a complexity of $\mathcal{O}((K+1)^{3.5})$ if an interior-point method is used, where $\mathcal{O}$ denotes the Landau notation. Thus, the computational complexity of $J$ randomized trials is on the order of $\mathcal{O}(J(K+1)^2)$, and the overall complexity of the proposed algorithm is $\mathcal{O}(L_1(K+1)^{3.5})+\mathcal{O}(L_1J(K+1)^2)$.

\subsection{Numerical Results}
In this section, we examine the performance of ARIS in interference mitigation for D2D communication scenario. A similar Rayleigh fading channel models are assumed for $\widetilde{\Hbf}$, $\widetilde\Gbf$, and $\widetilde\Dbf$ with elements whose variances are given by $\sigma_\text{h}^2$, $\sigma_\text{g}^2$, and $\sigma_\text{d}^2$, respectively. In the simulation, we set the noise variance as $\sigma^2=1$ and $L=6$. All the communication links have identical transmission power, i.e., $p_\ell=P$.

\begin{figure}
\centering
\includegraphics[width=3.3in]{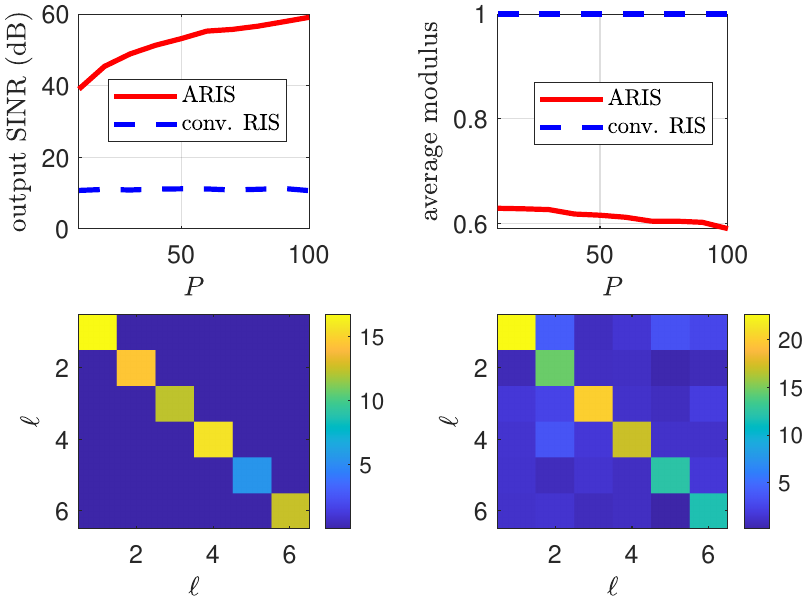}
\caption{Performance of ARIS in interference suppression for D2D communications when $\sigma^2_\text{h}=\sigma^2_\text{g}=\sigma_\text{d}^2=0$ dB and $K=64$. (a) Output SINR versus $P$; (b) Average ARIS modulus value ($\rho_k$) versus $P$; (c) Modulus of channel elements designed with ARIS; (d) Modulus of channel designed with conventional RIS.}
\label{fgr:results_device_to_device_P}
\end{figure}

\begin{figure}
\centering
\includegraphics[width=3.3in]{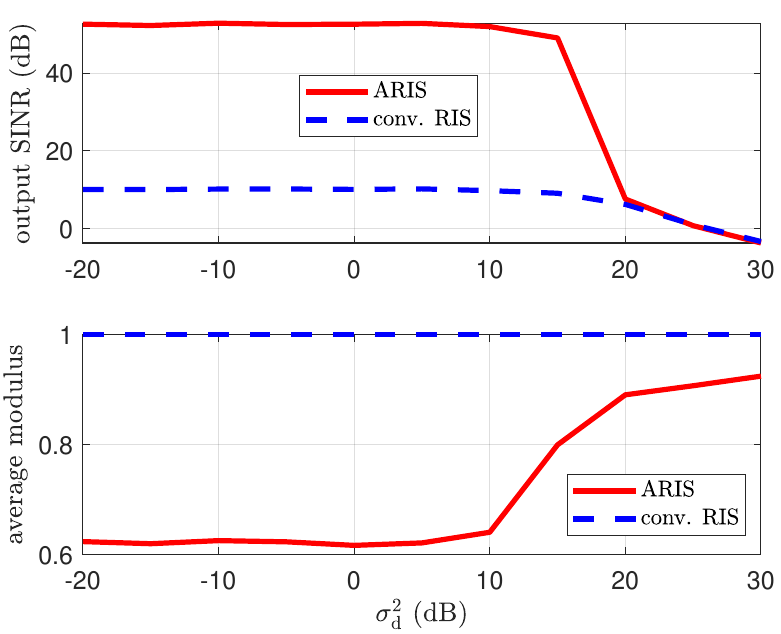}
\caption{Performance of RIS in interference suppression for device-to-device communications for various $\sigma_\text{d}^2$ when $\sigma^2_\text{h}=\sigma^2_\text{g}=0$ dB, $P=50$, and $K=64$. Top: Output SINR versus $\sigma_\text{d}^2$; Bottom: Average RIS modulus value ($\rho_k$) versus $\sigma_\text{d}^2$.}
\label{fgr:results_device_to_device_sigma_D}
\end{figure}

We first evaluate the performance when the direct-path channel strength between the TXs and RXs is small. Fig.\,\ref{fgr:results_device_to_device_P} (a) shows the output SINR versus the transmit power $P$ when $\sigma_\text{d}^2=0$ dB. It is seen that the output SINR of the design using the ARIS increases proportionally with the transmit power since the mutual interference channel can be completely cancelled using the ARIS, and the transmit power level does not affect the interference term. Unlike the case of radar and communication coexistence where it is desirable to zero out all the elements of the channel matrix between the BS and radar RX, here only the off-diagonal elements of the channel matrix $\widetilde{\Hbf}\Thetabf\widetilde{\Gbf}+\widetilde{\Dbf}$ are forced to zero since they account for the mutual interference among the D2D links. This is shown in Fig.\,\ref{fgr:results_device_to_device_P} (c) and (d), where the modulus of the channel elements between the TXs and RXs is plotted. In this case the ARIS achieves a diagonal channel matrix while the conventional RIS design cannot. The ability of the ARIS to cancel the mutual interference will be maintained as long as the direct channel strength $\sigma_d^2$ does not grow too large. This is observed in Fig.\,\ref{fgr:results_device_to_device_sigma_D}, where we observe that $\sigma_d^2$ must increase to nearly $20$ dB before it is necessary to reflect most of the received energy to try and cancel the interference. This is equivalent to the situation depicted in Fig.~\ref{fgr:toyexample}(d) where the reflected channel is not strong enough to overcome the direct channel component.

\begin{figure}
\centering
\includegraphics[width=3.3in]{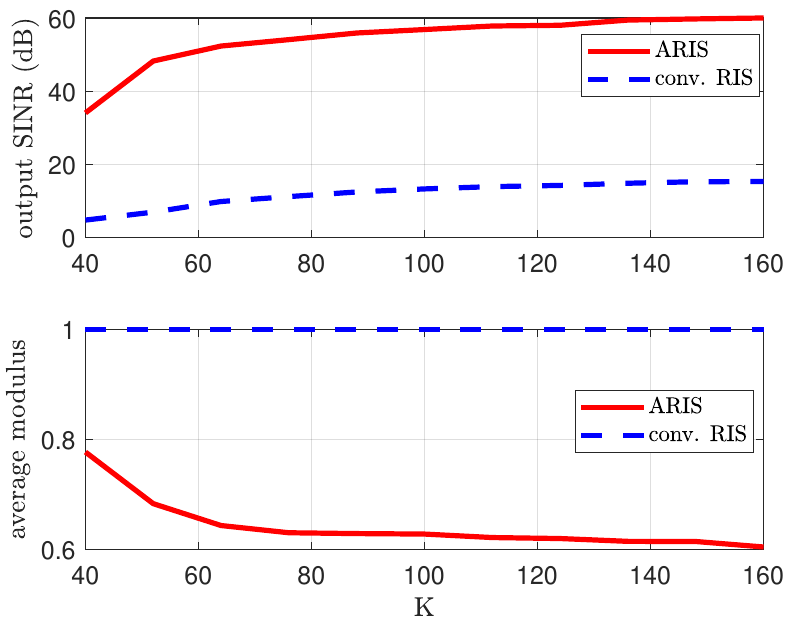}
\caption{Performance of RIS in interference suppression for device-to-device communications under various number of RIS elements $K$ when $P=50$, $\sigma^2_\text{h}=\sigma^2_\text{g}=0$ dB, and $\sigma^2_\text{d}=10$ dB. Top: Output SINR versus $K$; Bottom: Average RIS modulus value ($\rho_k$) versus $K$.}
\label{fgr:results_device_to_device_K}
\end{figure}

Finally, Fig.\,\ref{fgr:results_device_to_device_K} shows the output SINR versus the number of RIS elements $K$ when $P=50$, $\sigma^2_\text{h}=\sigma^2_\text{g}=0$ dB, and $\sigma^2_\text{d}=10$ dB. It is seen that the performance of both the ARIS and conventional RIS improves when the number of RIS elements $K$ increases. This is because a larger $K$ means more DoFs for the RIS to remove the mutual interference and enhance the desired communication links. However, the ARIS outperforms the conventional RIS for all $K$ due to its additional DoFs.

\section{ARIS for Physical Layer Security}
\label{sec:PLS}
In this section, we investigate the use of ARIS to enhance the physical layer security of a downlink communication system. We propose to maximize the secrecy rate to optimize the system security, and the resulting optimization problem is solved through fractional programming algorithm along with a sequential convex relaxation procedure.

\subsection{Problem Formulation}
\begin{figure}[tbh!]
\centering
\includegraphics[width=3.3in]{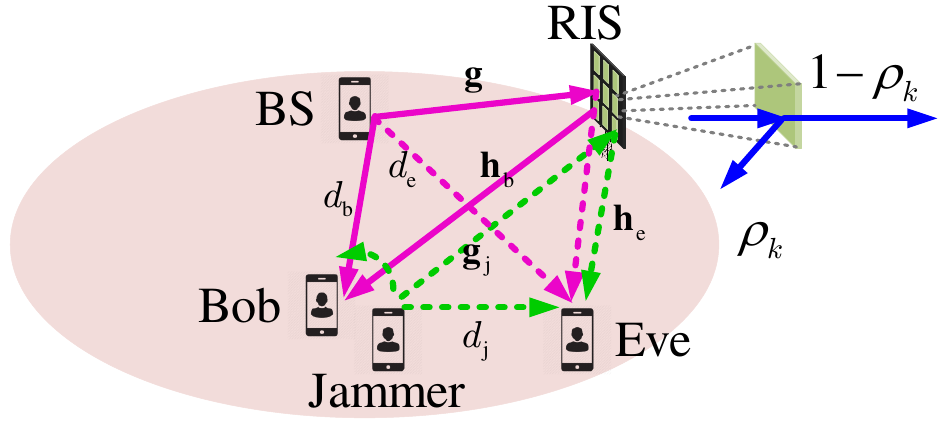}
\caption{ARIS-Assisted wiretap communication system.}
\label{fgr:PLSconfiguration}
\end{figure}
Consider an ARIS-assisted wiretap communication system, which consists of one BS (Alice), one legitimate user (Bob), one eavesdropper (Eve), and one ARIS, as illustrated in Fig.\,\ref{fgr:PLSconfiguration}. The BS, Bob, and Eve are all equipped with a single antenna. Meanwhile, an additional antenna is deployed near Bob for the transmission of a jamming signal. The received signal at Bob is given as
\ben\label{eq:received_bob}
y_\text{b}=(d_\text{b}+\gbf^T\Phibf\hbf_\text{b})x+d_\text{jb}\bar x+n_\text{b},
\een
where $d_\text{b}$ denotes the channel coefficient from the BS to Bob, $\gbf$ is the channel vector from the BS to the ARIS,  $\hbf_\text{b}$ is the channel vector from the ARIS to Bob, $x\sim\mathcal{CN}(0,1)$ is the transmitted symbol from the BS, $d_\text{jb}$ is the channel coefficient between the Jammer and Bob, $\bar x\sim\mathcal{CN}(0,1)$ is the transmitted symbol from the Jammer, and $n_\text{b}\sim\mathcal{CN}(0,\widetilde\sigma^2_\text{b})$ denotes the noise at Bob. Note that here we assume that most of the jamming signal is removed at Bob by employing successive interference cancellation since the transmitted jamming signal can be assumed to be known to Bob. Nonetheless, there is a residual self-interference term that we represent by $d_\text{jb}\bar x$. Thus, the received SINR can be expressed as
\ben\label{eq:SNR_Bob}
\text{SINR}_\text{b}=\frac{\vert d_\text{b}+\gbf^T\Phibf\hbf_\text{b}\vert^2}{\vert d_\text{jb}\vert^2+\widetilde\sigma^2_\text{b}}=\frac{\vert d_\text{b}+\gbf^T\Phibf\hbf_\text{b}\vert^2}{\sigma^2_\text{b}},
\een
where $\sigma^2_\text{b}\triangleq\vert d_\text{jb}\vert^2+\widetilde\sigma^2_\text{b}$.

Similarly, the received signal at Eve can be expressed as
\ben\label{eq:received_eve}
y_\text{e}=(d_\text{e}+\gbf^T\Phibf\hbf_\text{e})x+(d_\text{j}+\gbf_\text{j}^T\Phibf\hbf_\text{e})\bar x+n_\text{e},
\een
where $d_\text{e}$ is the channel between the BS and Eve, $\hbf_\text{e}$ the channel between the ARIS and Eve, $d_\text{j}$ the channel between the Jammer and Eve, $\gbf_\text{j}$ the channel between the Jammer and ARIS, and $n_\text{e}\sim\mathcal{CN}(0,\sigma^2_\text{e})$ denotes the noise at Eve. The corresponding received SINR at Eve is thus
\ben\label{eq:SINR_Eve}
\text{SINR}_\text{e}=\frac{\vert d_\text{e}+\gbf^T\Phibf\hbf_\text{e}\vert^2}{\vert d_\text{j}+\gbf_\text{j}^T\Phibf\hbf_\text{e}\vert^2+\sigma^2_\text{e}}.
\een
The secrecy rate is defined as the maximum rate of communication in which the information can be decoded arbitrarily and reliably at the legitimate receiver but not at the eavesdropper, i.e.,
\begin{align}\label{eq:SC}
\gamma(\Phibf)&=\Big[\log_2(1+\text{SINR}_\text{b})-\log_2(1+\text{SINR}_\text{e})\Big]^+\notag\\
&=\Big[\log_2\left(\frac{1+\text{SINR}_\text{b}}{1+\text{SINR}_\text{e}}\right)\Big]^+,
\end{align}
where $[\kappa]^+=\max(0,\kappa)$. In the following we will omit the $[\cdot]^+$ operation since we will be focusing on scenarios with a strictly positive secrecy rate.

The physical layer security problem of interest here is formulated as maximizing the secrecy rate by designing the ARIS $\Phibf$:
\begin{subequations}\label{eq:PLS}
\begin{gather}
\max_{\{\rho_k,\theta_k\}}~\frac{1+\text{SINR}_\text{b}}{1+\text{SINR}_\text{e}}\label{eq:PLS_cost}\\
\text{s.t.}~\eqref{eq:radarcomm_cons_1},~ \eqref{eq:radarcomm_cons_2}.\label{eq:PLS_const}
\end{gather}
\end{subequations}
Substituting \eqref{eq:SNR_Bob} and \eqref{eq:SINR_Eve} into \eqref{eq:PLS} and performing similar mathematical manipulations as in \eqref{eq:f}, the optimization problem in \eqref{eq:PLS} can be rewritten as
\begin{subequations}\label{eq:PLS_P1}
\begin{gather}
\max_{\bar\phibf}~\frac{1+\frac{\bar{\phibf}^H\Fbf_\text{b}\bar{\phibf}}{\sigma^2_\text{b}}}{1+\frac{\bar{\phibf}^H\Fbf_\text{e}\bar{\phibf}}{\bar{\phibf}^H\Fbf_\text{j}\bar{\phibf}+\sigma^2_\text{e}}}\\
\text{s.t.}~\vert\bar\phibf(k)\vert\leq 1, k=1,\cdots,K,~\text{and}~\vert\bar\phibf(K+1)\vert= 1\label{eq:PLS_P1_const},
\end{gather}
\end{subequations}
where
\begin{equation}
\begin{split}
\Fbf_\text{b}=
\begin{bmatrix}
\fbf_\text{b}\fbf_\text{b}^H& \fbf_\text{b} d_\text{b}\\
\fbf_\text{b}^H d_\text{b}^\ast& \vert d_\text{b}\vert^2
\end{bmatrix}, \quad
\Fbf_\text{e}=
\begin{bmatrix}
\fbf_\text{e}\fbf_\text{e}^H& \fbf_\text{e} d_\text{e}\\
\fbf_\text{e}^H d_\text{e}^\ast& \vert d_\text{e}\vert^2
\end{bmatrix},
\end{split}
\end{equation}
and
\begin{equation}
\begin{split}
\Fbf_\text{j}=
\begin{bmatrix}
\fbf_\text{j}\fbf_\text{j}^H& \fbf_\text{j} d_\text{j}\\
\fbf_\text{j}^H d_\text{j}^\ast& \vert d_\text{j}\vert^2
\end{bmatrix},
\end{split}
\end{equation}
with $\fbf_\text{b}^{\ast}=\gbf\circ\hbf_\text{b}$, $\fbf_\text{e}^{\ast}=\gbf\circ\hbf_\text{e}$, and $\fbf_\text{j}^{\ast}=\gbf_\text{j}\circ\hbf_\text{e}$.

Problem \eqref{eq:PLS_P1} can be further simplified as
\begin{subequations}\label{eq:PLS_P2}
\begin{gather}
\max_{\bar\phibf}~\frac{(\bar{\phibf}^H\Fbf_\text{b}\bar{\phibf}+\sigma^2_\text{b})(\bar{\phibf}^H\Fbf_\text{j}\bar{\phibf}+\sigma^2_\text{e})}{\bar{\phibf}^H\Fbf_\text{j}\bar{\phibf}+\bar{\phibf}^H\Fbf_\text{e}\bar{\phibf}+\sigma^2_\text{e}}\\
\text{s.t.}~\vert\bar\phibf(k)\vert\leq 1, k=1,\cdots,K,~\text{and}~\vert\bar\phibf(K+1)\vert= 1\label{eq:PLS_P2_const} .
\end{gather}
\end{subequations}
The SDR form of \eqref{eq:PLS_P2} is
\begin{subequations}\label{eq:PLS_SDR}
\begin{gather}
\max_{\bar\Phibf}\,\frac{\sigma^2_\text{b}\sigma^2_\text{e}+\sigma^2_\text{e}\text{tr}\big(\bar{\Phibf}\Fbf_\text{b}\big)+\sigma^2_\text{b}\text{tr}(\bar{\Phibf}\Fbf_\text{j})+\text{tr}\big(\bar{\Phibf}\Fbf_\text{b}\bar{\Phibf}\Fbf_\text{j}\big)}{\sigma^2_\text{e}+\text{tr}\big(\bar{\Phibf}(\Fbf_\text{e}+\Fbf_\text{j})\big)}\\
\text{s.t.}\,\vert\bar\Phibf(k,k)\vert\leq 1, k=1,\cdots,K,\label{eq:PLS_SDR_const1}\\
\vert\bar\Phibf(K+1,K+1)\vert= 1.\label{eq:PLS_SDR_const2}
\end{gather}
\end{subequations}
The above fractional programming problem can be solved through the Dinkebach algorithm by introducing a slack variable $\lambda$, in which case problem \eqref{eq:PLS_SDR} becomes
\begin{subequations}\label{eq:PLS_SDR_Dinkelback}
\begin{gather}
\max_{\bar\Phibf,\lambda}\,\gamma(\bar\Phibf,\lambda)\\
\text{s.t.}\,\eqref{eq:PLS_SDR_const1},\eqref{eq:PLS_SDR_const2},
\end{gather}
\end{subequations}
where
\begin{align}
&\gamma(\bar\Phibf,\lambda)=\sigma^2_\text{b}\sigma^2_\text{e}+\sigma^2_\text{e}\text{tr}\big(\bar{\Phibf}\Fbf_\text{b}\big)+\sigma^2_\text{b}\text{tr}(\bar{\Phibf}\Fbf_\text{j})\notag\\
&+\text{tr}\big(\bar{\Phibf}\Fbf_\text{b}\bar{\Phibf}\Fbf_\text{j}\big)-\lambda\Big(\sigma^2_\text{e}+\text{tr}\big(\bar{\Phibf}(\Fbf_\text{e}+\Fbf_\text{j})\big)\Big).
\end{align}

When we fix $\bar\Phibf$ to the value obtained from the $i$-th iteration $\bar\Phibf^{(i)}$, the $\lambda^{(i+1)}$ can be obtained in closed form as
\begin{align}\label{eq:lambdacomputing}
&\lambda^{(i+1)}\notag\\
&=\frac{\sigma^2_\text{b}\sigma^2_\text{e}+\sigma^2_\text{e}\text{tr}(\bar{\Phibf}^{(i)}\Fbf_\text{b})+\sigma^2_\text{b}\text{tr}(\bar{\Phibf}^{(i)}\Fbf_\text{j})+\text{tr}\big(\bar{\Phibf}^{(i)}\Fbf_\text{b}\bar{\Phibf}^{(i)}\Fbf_\text{j}\big)}{\sigma^2_\text{e}+\text{tr}\big(\bar{\Phibf}^{(i)}(\Fbf_\text{e}+\Fbf_\text{j})\big)}.
\end{align}
Next, we can find $\bar\Phibf^{(i+1)}$ with fixed $\lambda^{(i+1)}$ by solving the following optimization problem
\begin{subequations}\label{eq:PLS_P1_SDR_Dinkelback}
\begin{gather}
\max_{\bar\Phibf,\lambda}\,\gamma(\bar\Phibf,\lambda^{(i+1)})\\
\text{s.t.}\,\eqref{eq:PLS_SDR_const1},\eqref{eq:PLS_SDR_const2},
\end{gather}
\end{subequations}
which still is a nonconvex problem due to the fact we are maximizing the quadratic term $\text{tr}\big(\bar{\Phibf}\Fbf_\text{b}\bar{\Phibf}\Fbf_\text{j}\big)$ w.r.t. $\bar\Phibf$. We can employ a convex relaxation based sequential convex programming (SCP) approach to solve the problem. Specifically, let $\bar\Phibf^{(i_1)}$ denote the solution from the $i_1$-th iteration of the SCP iterative process. Then, $\text{tr}\big(\bar{\Phibf}\Fbf_\text{b}\bar{\Phibf}\Fbf_\text{j}\big)$ can be approximated by its first-order Taylor expansion at $\bar\Phibf^{(i_1)}$ as
\begin{align}
&\text{tr}\big(\bar{\Phibf}\Fbf_\text{b}\bar{\Phibf}\Fbf_\text{j}\big)\approx\text{tr}\big(\bar{\Phibf}^{(i_1)}\Fbf_\text{b}\bar{\Phibf}^{(i_1)}\Fbf_\text{j}\big)\notag\\
&+\text{tr}\Big(\big(\Fbf_\text{b}\bar{\Phibf}^{(i_1)}\Fbf_\text{j}+\Fbf_\text{j}\bar{\Phibf}^{(i_1)}\Fbf_\text{b}\big)\big(\bar{\Phibf}-\bar{\Phibf}^{(i_1)}\big)\Big).
\end{align}
Therefore, we can solve the following convex optimization problem
\begin{subequations}\label{eq:PLS_P2_SDR_Dinkelback}
\begin{gather}
\max_{\bar\Phibf}\,\widetilde\gamma(\bar\Phibf,\bar\Phibf^{(i_1)},\lambda^{(i+1)})\\
\text{s.t.}\,\eqref{eq:PLS_SDR_const1},\eqref{eq:PLS_SDR_const2},
\end{gather}
\end{subequations}
where
\begin{align}
&\widetilde\gamma(\bar\Phibf,\bar\Phibf^{(i_1)},\lambda^{(i+1)})=\sigma^2_\text{b}\sigma^2_\text{e}+\sigma^2_\text{e}\text{tr}\big(\bar{\Phibf}\Fbf_\text{b}\big)+\sigma^2_\text{b}\text{tr}(\bar{\Phibf}\Fbf_\text{j})\notag\\
&+\text{tr}\big(\bar{\Phibf}^{(i_1)}\Fbf_\text{b}\bar{\Phibf}^{(i_1)}\Fbf_\text{j}\big)\notag\\
&+\text{tr}\Big(\big(\Fbf_\text{b}\bar{\Phibf}^{(i_1)}\Fbf_\text{j}+\Fbf_\text{j}\bar{\Phibf}^{(i_1)}\Fbf_\text{b}\big)\big(\bar{\Phibf}-\bar{\Phibf}^{(i_1)}\big)\Big)\notag\\
&-\lambda^{(i+1)}\Big(\sigma^2_\text{b}\sigma^2_\text{e}+\sigma^2_\text{b}\text{tr}\big(\bar{\Phibf}(\Fbf_\text{e}+\Fbf_\text{j})\big)\Big).
\end{align}
The SCP iterations end when the difference between the cost function of two successive iterations is smaller than some tolerance. Then, the obtained $\bar\Phibf^{(i_1)}$ is utilized as $\bar\Phibf$ for the next Dinkebach iteration, i.e., $\bar\Phibf^{(i+1)}$.

The outer iteration of the Dinkebach algorithm is repeated until the algorithm converges, e.g., the secrecy improvement is smaller than a given tolerance. Our proposed solution is summarized in Algorithm \ref{alg:PLS}.

Similarly, the formulation for the design using the standard RIS can be obtained by changing the inequality constraint in \eqref{eq:PLS_P1_const} to a equality one, and the solution can be accordingly adjusted with the new constraint without imposing additional difficulties.

\begin{algorithm}
\caption{Sequential Convex Relaxation Based Dinkebach algorithm to Solve \eqref{eq:PLS}}
\begin{algorithmic}
\label{alg:PLS}
\STATE \textbf{input:} All channel parameters defined in \eqref{eq:received_bob} and \eqref{eq:received_eve}.
\STATE \textbf{output:} ARIS design parameters $\Phibf$.
\STATE  \textbf{initialization:} Set iteration index $i=0$ and initialize $\phibf$ as $\phibf^{(0)}$.
\\
\REPEAT
\STATE
\begin{enumerate}
\item Use $\bar\Phibf^{(i)}$ to Compute $\lambda^{(i+1)}$ through \eqref{eq:lambdacomputing}.
\item Set inner iteration index $i_1=0$ and $\bar\Phibf^{(i_1)}=\bar\Phibf^{(i)}$
\STATE \textbf{repeat}
\begin{enumerate}
\item Fix $\bar\Phibf^{(i_1)}$ and $\lambda^{(i+1)}$.
\item Update $\bar\Phibf^{(i_1+1)}$ by solving \eqref{eq:PLS_P2_SDR_Dinkelback}.
\item Set $i_1=i_1+1$.
\end{enumerate}
\textbf{until} convergence

\item Let $\bar\Phibf^{(i+1)}=\bar\Phibf^{(i_1+1)}$.
\item Set $i=i+1$
\end{enumerate}
\UNTIL convergence.
\end{algorithmic}
\end{algorithm}

\subsection{Numerical Results}
In the simulations that follow, $d_\text{b}$, $\gbf$, $\hbf_\text{b}$, $d_\text{e}$, $\hbf_\text{e}$, $d_\text{j}$, and $\gbf_\text{j}$ are all Rayleigh fading channels with channel strengths $\sigma_\text{db}^2$, $\sigma_\text{g}^2$, $\sigma_\text{hb}^2$, $\sigma_\text{de}^2$, $\sigma_\text{he}^2$, $\sigma_\text{dj}^2$, and $\sigma_\text{gj}^2$, respectively. The noise variance $\sigma^2_\text{e}=\sigma^2_\text{b}=1$.

\begin{figure}[tbh!]
\centering
\includegraphics[width=3.3in]{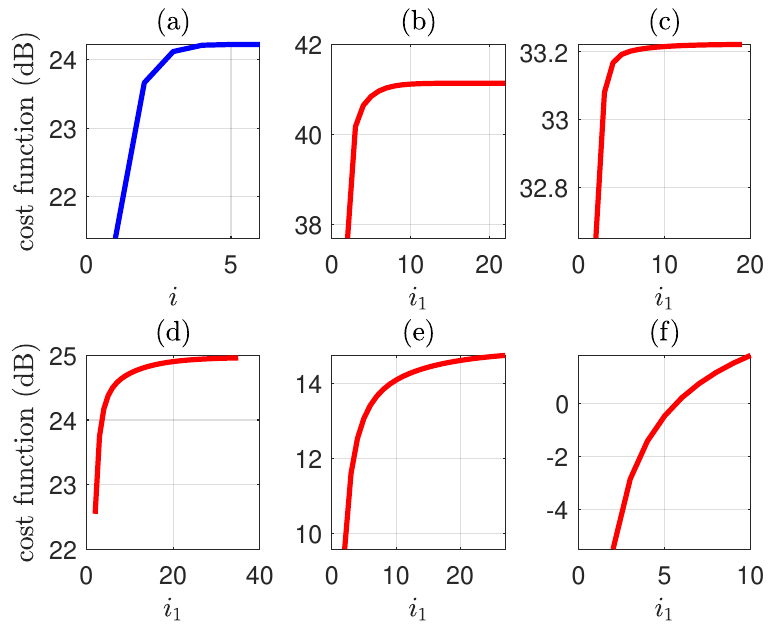}
\caption{Convergence behaviour of Algorithm \ref{alg:PLS} when $K=16$ and all the channel strengths are 0 dB. (a) Outer iterations; (b)-(f): the $1$st to the $5$th inner iterations.}
\label{fgr:PLSconvergence}
\end{figure}

We first demonstrate the convergence of the proposed algorithms. Fig.\,\ref{fgr:PLSconvergence} depicts the output cost function value in \eqref{eq:PLS_cost} of the ARIS design versus the number of iterations when $K=16$ and all the channel strengths are $0$ dB. Fig.\,\ref{fgr:PLSconvergence} (a) is the outer iterations of Algorithm \ref{alg:PLS} and it can be seen that the Dinkebach algorithm converges in $5$ iterations. Fig.\,\ref{fgr:PLSconvergence} (b)-(f) are the $5$ corresponding inner iterations and the results show that the SCP inner iterations also converge quickly.

\begin{figure}[tbh!]
\centering
\includegraphics[width=3.3in]{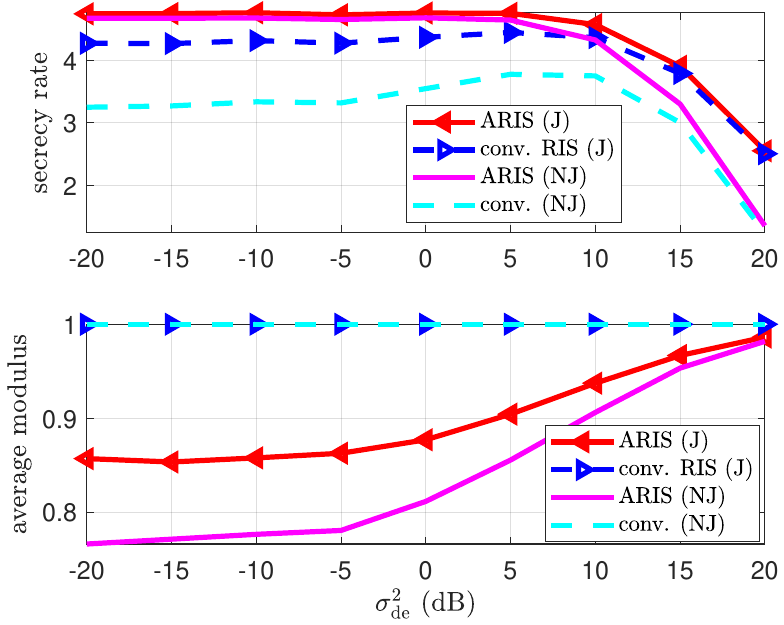}
\caption{Performance of ARIS for PLS under various direct-path channel strength between the BS and Eve. Top: Secrecy rate versus $\sigma_{\text{de}}^2$, bottom: modulus of ARIS elements versus $\sigma_{\text{de}}^2$.}
\label{fgr:PLSvarioussigmade}
\end{figure}

Next, we evaluate the effect of the channel strengths by considering four scenarios: Using ARIS and standard RIS without deploying a Jammer, denoted as \textbf{ARIS (NJ)} and \textbf{conv. RIS (NJ)}, respectively; and using ARIS and standard RIS with a Jammer, denoted as \textbf{ARIS (J)} and \textbf{conv. RIS (J)}. Fig.\,\ref{fgr:PLSvarioussigmade} shows the output secrecy rate $\gamma(\Phibf)$ and the corresponding average ARIS modulus versus the direct-path channel strength between the BS and Eve, i.e., $\sigma_\text{de}^2$, where $\sigma_\text{dj}^2=\sigma_\text{gj}^2=\sigma_\text{he}^2=\sigma_\text{hb}^2=0$ dB, $\sigma_\text{db}^2=\sigma_\text{g}^2=10$ dB, and $K=2$. It is seen that for the both the scenarios with and without a Jammer, the ARIS outperforms the conventional RIS for small $\sigma_\text{de}^2$ since in such cases the ARIS can completely cancel the direct-path transmission between the BS and Eve by adjusting the phase and modulus of the RIS elements, which in turn removes information leakage to Eve. This cannot be achieved by only adjusting the phase of the RIS elements. In addition, for small $\sigma_\text{de}^2$ the Jammer provides no benefit since the ARIS by itself can guarantee no information leakage. As the direct-path channel strength $\sigma_\text{de}^2$ increases, the benefit of absorption disappears since the ARIS needs to maximize its reflected energy to cancel as much of the direct-path transmission as possible. Since the overall transmission between the BS and Eve cannot be completely removed by the ARIS when $\sigma_\text{de}^2$ increases, the performance in all scenarios degrades. In such cases, the system benefits from the presence of the Jammer, as verified by the performance gap at high $\sigma_\text{de}^2$ in Fig.\,\ref{fgr:PLSvarioussigmade}.

\begin{figure}[tbh!]
\centering
\includegraphics[width=3.3in]{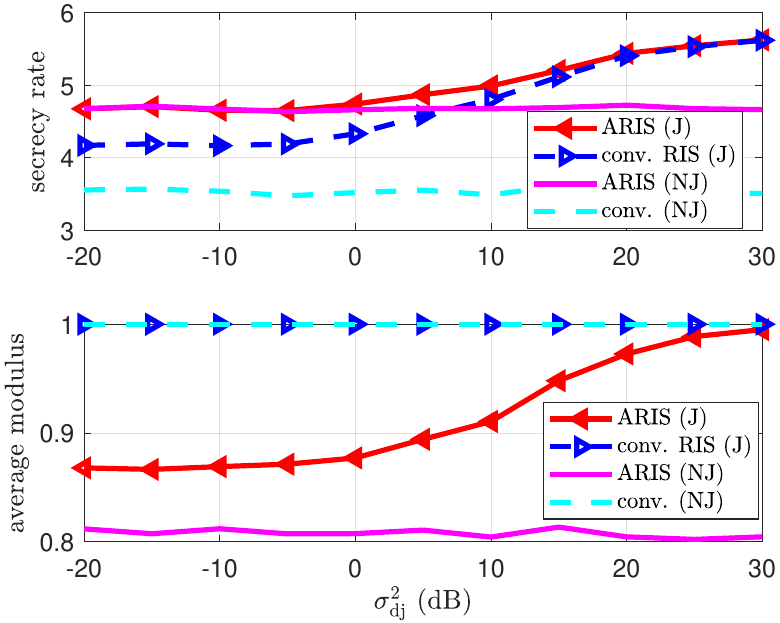}
\caption{Performance of ARIS in the PLS application under various strengths of the direct channel between the Jammer and Eve, i.e., $\sigma_{\text{dj}}^2$.}
\label{fgr:PLSvarioussigmadj}
\end{figure}

Fig.\,\ref{fgr:PLSvarioussigmadj} depicts the performance of the proposed approach under various strengths of the direct-path channel between the Jammer and Eve when $\sigma_\text{de}^2=\sigma_\text{gj}^2=\sigma_\text{he}^2=\sigma_\text{hb}^2=0$ dB, $\sigma_\text{db}^2=\sigma_\text{g}^2=10$ dB, and $K=2$. For a weak Jammer channel, the ARIS is the key to providing increased secrecy compared with the case of a conventional RIS As the strength of the Jammer channel increases, the impact of the Jammer becomes larger than that of the RIS, and the ARIS and conventional RIS converge to the same performance.



\section{Conclusion}
\label{sec:conclusion}
We have studied the benefit of an ARIS that possess the capability of adjusting both the reflection coefficient phase and gain through absorption. We presented three applications where the ARIS can achieve substantial performance gains compared with conventional RIS with phase-only control: coexistence of radar and communication systems, device-to-device communications, and jammer-aided physical layer security. These examples demonstrate that the extra degrees of freedom provided by the adaptive ARIS absorption are especially beneficial for situations where interference suppression is the key to achieving desired system performance. Future directions of interest related to ARIS include extensions that consider realistic models for the achievable RIS amplitude and phase, consideration of mutual interference where the ARIS is deployed to assist both the communication and radar functionality in ISAC systems, and study of the impact of imperfect CSI on ARIS relative to conventional RIS.
\bibliographystyle{IEEEtran}
\bibliography{Hybrid_RIS}
\end{document}

%% file: common.tex
\newcommand{\ba}{\begin{array}}
\newcommand{\ea}{\end{array}}
\newcommand{\be}{\begin{displaymath}}
\newcommand{\ee}{\end{displaymath}}
\newcommand{\ben}{\begin{equation}}
\newcommand{\een}{\end{equation}}
\newcommand{\bena}{\begin{eqnarray}}
\newcommand{\eena}{\end{eqnarray}}
\newcommand{\beqa}{\begin{eqnarray*}}
\newcommand{\enqa}{\end{eqnarray*}}

\newcommand{\bc}{\begin{center}}
\newcommand{\ec}{\end{center}}
\newcommand{\bi}{\begin{itemize}}
\newcommand{\ei}{\end{itemize}}
\newcommand{\benu}{\begin{enumerate}}
\newcommand{\eenu}{\end{enumerate}}
\newcommand{\bdes}{\begin{description}}
\newcommand{\edes}{\end{description}}
\newcommand{\bt}{\begin{tabular}}
\newcommand{\et}{\end{tabular}}

\newcommand \Phibf{\boldsymbol{\Phi}}

\newcommand \xibf{\boldsymbol{\xi}}

\newcommand \phibf{\boldsymbol{\phi}}

\newcommand \Thetabf{\boldsymbol{\Theta}}

\newcommand \Xibf{\boldsymbol{\Xi}}

\newcommand \abf{{\bf a}}

\newcommand \dbf{{\bf d}}

\newcommand \fbf{{\bf f}}
\newcommand \gbf{{\bf g}}
\newcommand \hbf{{\bf h}}

\newcommand \ubf{{\bf u}}

\newcommand \Abf{{\bf A}}

\newcommand \Dbf{{\bf D}}

\newcommand \Fbf{{\bf F}}
\newcommand \Gbf{{\bf G}}
\newcommand \Hbf{{\bf H}}

\newcommand \Xbf{{\bf X}}
\newcommand \Ybf{{\bf Y}}

\newcommand{\circlambda}{\mbox{$\Lambda$
             \kern-.85em\raise1.5ex
             \hbox{$\scriptstyle{\circ}$}}\,}

